\documentclass[twocolumn]{article}

\usepackage[dvips]{epsfig}

\def\thebibliography#1{\section*{\normalsize \bf References 
 }\list
 {[\arabic{enumi}]}{\settowidth\labelwidth{[#1]}\leftmargin\labelwidth
 \advance\leftmargin\labelsep
 \usecounter{enumi}}
 \def\newblock{\hskip .11em plus .33em minus .07em}
 \sloppy\clubpenalty4000\widowpenalty4000
 \sfcode`\.=1000\relax}

\begin{document}

\twocolumn[

\mbox{}
\vspace{50mm}

\begin{center} \LARGE 
   Effective mass at the surface of a Fermi liquid \\
\end{center}

\begin{center} \large
   M. Potthoff and W. Nolting
\end{center}

\begin{center} \small \it 
   Institut f\"ur Physik, 
   Humboldt-Universit\"at zu Berlin, 
   Germany
\end{center}
\vspace{10mm}

\small

---------------------------------------------------------------------------------------------------------------------------------------------------

{\bf Abstract} \\

Using the dynamical mean-field theory, we calculate the effective 
electron mass in the Hubbard model on a semi-infinite lattice. 
At the surface the effective mass is strongly enhanced. 
Near half-filling this gives rise to a correlation-driven 
one-electron surface mode.
\vspace{2mm}

---------------------------------------------------------------------------------------------------------------------------------------------------

\vspace{10mm} 
]

Important characteristics of interacting electrons on a lattice
can be studied within the Hubbard model. If symmetry-broken 
phases are ignored, the electron system is generally expected 
to form a Fermi-liquid state in dimensions $D>1$. The effective 
mass $m^\ast$ which determines the quasi-particle dispersion 
close to the Fermi energy, can be substantially enhanced for 
strong interaction $U$. A further enhancement may be possible at 
the surface of a $D=3$ lattice since here the reduced coordination 
number $z_s$ of the surface sites results in a reduced variance 
$\Delta \propto z_s$ of the free local density of states. This 
implies the effective interaction $U/\sqrt{\Delta}$ to be larger 
and thereby a tendency to strengthen correlation effects at the 
surface. Using the dynamical mean-field theory (DMFT) \cite{dmft}, 
we investigate the strongly correlated Hubbard model with uniform
nearest-neighbor hopping $t$ on a semi-infinite sc(100) lattice
at zero temperature. Close to half-filling, the surface effective
mass is found to be {\em strongly} enhanced compared with the bulk 
value.

We have generalized DMFT to film geometries \cite{PN98}. A film 
consisting of $d=15$ layers with the normal along the [100] 
direction turns out to be sufficient to simulate the actual sc(100) 
surface, i.~e.\ bulk properties are recovered at the film center. 
The lattice problem is mapped onto a set of $d$ single-impurity 
Anderson models (SIAM) which are indirectly coupled via the 
self-consistency condition of DMFT: $G_{0}^{(\alpha)}(E) = \left( 
G_{\alpha \alpha}(E)^{-1} + \Sigma_\alpha(E) \right)^{-1}$. 
The iterative solution starts with the layer-dependent local 
self-energy $\Sigma_\alpha$ ($\alpha=1,...,d$) which determines 
the layer-dependent on-site elements of the one-electron lattice 
Green function $G_{\alpha \alpha}$ via the (lattice) Dyson equation. 
The DMFT self-consistency equation then fixes the free impurity 
Green function $G_{0}^{(\alpha)}$ of the $\alpha$-th SIAM. 
Following the exact-diagonalization approach \cite{CK94}, we 
take a finite number of $n_s$ sites in the impurity model. 
The parameters of (the $\alpha$-th) SIAM are then obtained by 
fitting $G_{0}^{(\alpha)}$. Finally, the exact numerical solution 
of the $\alpha$-th SIAM yields a new estimate for $\Sigma_\alpha$
which is required for the next cycle. From the self-consistent 
solution the effective mass is calculated as $m_\alpha^\ast = 
1 - d \Sigma_\alpha(0) / dE$.

%++++++++++++++++++++++++++++++++++++++++++++++++++++++++++
\begin{figure}[t] 
\vspace{-3mm}
\center{\psfig{figure=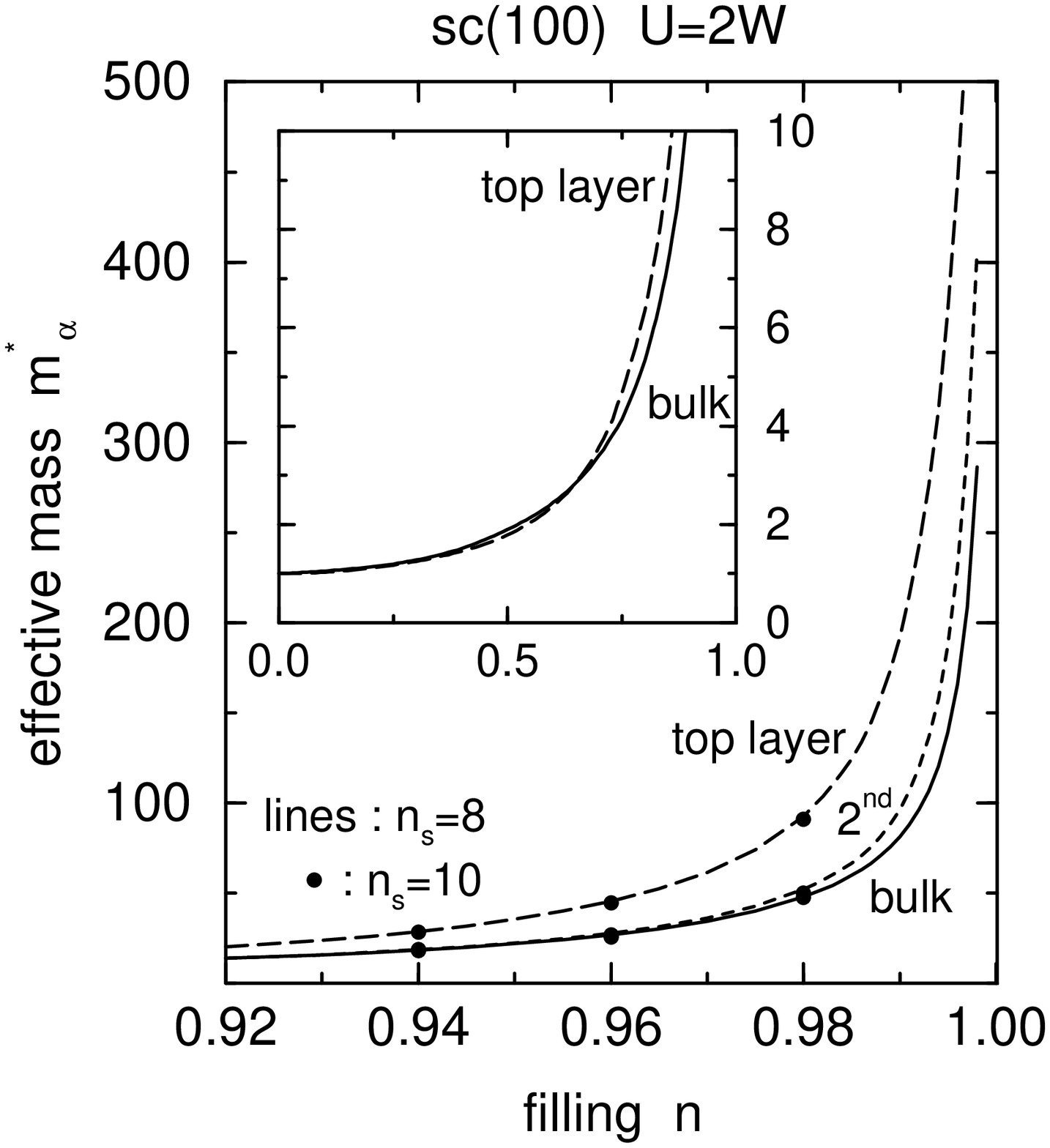,width=76mm,angle=0}}
\vspace{-5mm}

\parbox[]{76mm}{\small Fig.~1.
$m_\alpha^\ast$ for the top- and sub-surface layer and the bulk as 
a function of the filling $n$. $U=24 |t| = 2W$ ($W$ with of the free 
bulk density of states).
\label{fig:mass}
}
\end{figure}
%++++++++++++++++++++++++++++++++++++++++++++++++++++++++++

The figure shows the bulk mass $m_b^\ast$ to increase with 
increasing filling $n$. For $U=2W$ the system is a Mott-Hubbard 
insulator at half-filling ($n=1$). Consequently, $m_b^\ast$ diverges 
for $n \mapsto 1$. Up to $n \approx 0.98$ all layer-dependent masses 
$m_\alpha^\ast$ ($\alpha = 2,3,...$) are almost equal except for the 
top-layer mass $m_s^\ast$ ($\alpha=1$). For $n\mapsto 1$, $m_s^\ast$ 
is considerably enhanced, e.~g.\ $m_s^\ast/m_b^\ast=2.4$ for $n=0.99$ 
\cite{conv}.

There is an interesting consequence of this finding: Close to the 
Fermi energy where damping effects are unimportant, the Green 
function $G_{\alpha \beta}({\bf k}, E)$ can be obtained from the 
low-energy expansion of the self-energy: $\Sigma_\alpha(E) = 
a_\alpha + (1 - m_\alpha^\ast) E + \cdots$. For each wave vector 
${\bf k}$ of the two-dimensional surface Brillouin zone (SBZ), 
the poles of $G_{\alpha \beta}({\bf k}, E)$ yield the quasi-particle 
energies $\eta_r({\bf k})$ ($r=1,...,d$). The $\eta_r({\bf k})$ are 
the eigenvalues of the renormalized hopping matrix $T_{\alpha \beta}
({\bf k}) \equiv (m_\alpha m_\beta)^{-1/2} (\epsilon_{\alpha \beta}
({\bf k})+\delta_{\alpha\beta} (a_\alpha-\mu))$ where the non-zero 
elements of the free hopping matrix are given by $\epsilon_\|({\bf k}) 
\equiv \epsilon_{\alpha \alpha}({\bf k}) = 2t (\cos k_x + \cos k_y)$ 
and $\epsilon_\perp({\bf k}) \equiv \epsilon_{\alpha\alpha \pm 1}
({\bf k}) = |t|$. For the semi-infinite system ($d\mapsto \infty$) 
the eigenvalues form a continuum of (bulk) excitations at a given 
${\bf k}$. A surface excitation can split off the bulk continuum if 
$m_s^\ast / m_b^\ast$ is sufficiently large. Assuming $m_\alpha^\ast 
= m_b^\ast$ for $\alpha \ne 1$ and $m_{\alpha=1}^\ast = m_s^\ast$
(cf.\ Fig.~1), a simple analysis of $T_{\alpha\beta}({\bf k})$ yields 
the following analytical criterion for the existence of a surface 
mode in the low-energy (coherent) part of the one-electron spectrum 
($r^2 \equiv m_b^\ast/ m_s^\ast < 1$):
\begin{eqnarray}
  \frac{2-r^2}{1-r^2} <
  \left| 
  \frac{\epsilon_\|({\bf k}) - \mu + a_s}{\epsilon_\perp({\bf k})}
  + \frac{a_b - a_s}{(1-r^2) \epsilon_\perp({\bf k})}
  \right| \: .
\nonumber
\end{eqnarray}
The evaluation for the present case shows that indeed a surface mode
can split off at the $(0,0)$ and the $(\pi,\pi)$ points in the
SBZ for fillings $n>0.94$ and $n>0.90$, respectively. 
The surface modification of 
$a_\alpha=\Sigma_\alpha(0)$ (second term on the r.h.s) 
turns out to be unimportant for the effect.
The excitation found is thus a correlation-driven surface excitation 
which is caused by the strong specific surface renormalization of the 
effective mass.
\newline
\newline


\begin{thebibliography}{10}
\bibitem{conv}
The systematic comparison of the results for $n_s=8$ with those for
 $n_s=6$ and (at a few points) also with $n_s=10$ (Fig.~1) shows
 finite-size effects to be unimportant up to $n=0.98$ at least.
 We are thus confident to have the essentially exact solution of 
 the mean-field equations.
\end{thebibliography}
\end{document}